\documentclass[prl, showpacs, showkeys, twocolumn]{revtex4}
\usepackage{bm}
\usepackage{amssymb, amsfonts, amsbsy}
\begin{document}
%------------------------------------------------
\def\TODAY{12 March 2009; 17 March 2009}
%------------------------------------------------
\title{Birkhoff-like theorem for rotating stars in (2+1) dimensions}
%------------------------------------------------
\author{Jozef Sk\'akala}
\email{jozef.skakala@msor.vuw.ac.nz}
\affiliation{School of Mathematics, Statistics, and Operations Research,
Victoria University of Wellington, Wellington, New Zealand}
%------------------------------------------------
\author{Matt Visser}
\email{matt.visser@msor.vuw.ac.nz}
\affiliation{School of Mathematics, Statistics, and Operations Research,
Victoria University of Wellington, Wellington, New Zealand}
%------------------------------------------------
%------------------------------------------------
\date{\TODAY;  \LaTeX-ed \today}
%------------------------------------------------
\begin{abstract}
%------------------------------------------------
Consider a rotating and possibly pulsating ``star'' in (2+1) dimensions. If the star is axially symmetric, then in the vacuum region surrounding the star, (a region that we assume at most contains a cosmological constant), the Einstein equations imply that under physically plausible conditions the geometry is in fact stationary. Furthermore, the  geometry external to the star is then uniquely guaranteed to be the (2+1) dimensional analogue of the Kerr-de~Sitter spacetime, the BTZ geometry.  This Birkhoff-like theorem is very special to (2+1) dimensions, and fails in (3+1) dimensions. Effectively, this is a ``no hair'' theorem for (2+1) dimensional axially symmetric stars: the exterior geometry is completely specified by the mass, angular momentum, and cosmological constant. 

%------------------------------------------------
\end{abstract}
%------------------------------------------------
%-----------------------------------------------------------------
%-----------------------------------------------------------------
%-----------------------------------------------------------------
%\pacs{}
%------------------------------------------------
\maketitle
%------------------------------------------------
%------------------------------------------------
\def\d{{\mathrm{d}}}
\def\implies{\Rightarrow}
\newcommand{\scri}{\mathscr{I}}
\newcommand{\sun}{\ensuremath{\odot}}
%------------------------------------------------
%-----------------------------------------------------------------------------
%\def\lint{\hbox{\Large $\displaystyle\int$}} %needs \usepackage{amssymb}
%\def\hint{\hbox{\Huge $\displaystyle\int$}}  %needs \usepackage{amssymb}
%-----------------------------------------------------------------------------
\def\eg{{\it e.g.}}
\def\ie{{\it i.e.}}
\def\etc{{\it etc.}}
\def\sign{{\hbox{sign}}}
%-------------------------------------------------------------------------
\def\eof{\Box}
%-------------------------------------------------------------------------

\noindent\emph{Introduction:} For a spherically symmetric but possibly pulsating star, the Birkhoff theorem~\cite{Birkhoff, Jebsen, Deser, Ravndal} implies that the vacuum region surrounding the star can be proved to be static, and is actually a portion of the Schwarzschild (or Schwarzschild-de Sitter spacetime). While this theorem was first established in (3+1) dimensions, it was rapidly generalized to ($d$+1) dimensions. 

In contrast,  there is no Birkhoff theorem for rotating (axially symmetric) stars in (3+1) dimensions.
In (3+1) dimensions it is  \emph{not} true that the spacetime geometry in the vacuum region outside a generic rotating star (or planet)  is a part of the Kerr (or Kerr-de~Sitter) geometry~\cite{Kerr, Kerr-intro, Kerr0, Kerr1}.   The best result one can obtain is the much milder statement that outside a rotating star (or planet) the geometry asymptotically approaches the Kerr geometry.  The basic problem is that in the (3+1) dimensional Kerr geometry all the multipole  moments are very closely related to each other --- whereas in real  physical stars the mass quadrupole, octopole, and higher moments of the mass distribution can in principle be independently specified. Of course higher  $n$-pole fields fall off as $1/r^{2+n}$, so that far away from the object the lowest multipoles dominate --- it is in this \emph{asymptotic} sense that the  Kerr geometry is relevant for rotating stars.
  
On the other hand, in (2+1) dimensions, the Einstein equations are much simpler. Outside the star the spacetime is (locally) a constant curvature spacetime;  it is locally one of the de Sitter/ Minkowski/ anti-de~Sitter  spacetimes depending on the sign of the cosmological constant. This gives us some hope, borne out by direct calculation~\cite{AMZ}, that the situation might be more tractable there, and that a Birkhoff-like theorem may be achievable.  

In this article we shall present a physically motivated and completely  ``elementary'' derivation of a suitable Birkhoff-like theorem for (2+1) dimensions, focussing on physically plausible and appropriate coordinate choices, efficient use of the Einstein equations, and clear physical interpretation of the results. The Birkhoff-like results of~\cite{AMZ} and the current article may perhaps best be viewed as a ``no hair'' theorem for stars in (2+1) dimensions: The exterior geometry of an axially symmetric star is completely specified by the mass, angular momentum, and cosmological constant. 

%\enlargethispage{125pt}

%\bigskip

\noindent
%-----------------------------------------------------------------------------
\noindent\emph{Strategy:} 
The basic strategy we adopt will be this:\\
1) Use axial symmetry and coordinate transformations to reduce the (2+1) line element to a canonical form involving three independent functions of $r$ and $t$.\\
2) Use \emph{some} of the vacuum Einstein equations, in the form $R_{ab}=2\Lambda g_{ab}$, to \emph{deduce} that outside the pulsating rotating star the line element is in fact independent of $t$, so the exterior spacetime is \emph{proved} to be stationary.\\
3) Use a variant of the BTZ analysis~\cite{BTZ, BTZ2} to show that the exterior spacetime is a (2+1) version of Kerr-deSitter/ Kerr/ Kerr-anti-deSitter spacetime.

%\bigskip
%-----------------------------------------------------------------------------
\noindent\emph{Canonical form of the metric:} 
Axial symmetry implies that one can introduce a coordinate $\phi$ that runs from $0$ to $2\pi$ and such that the metric is independent of $\phi$. 
Let the other two coordinates be called $t$ and $r$, and let
\begin{eqnarray}
a,b,c\dots &\in&\{0,1,2\} \hbox{ corresponding to } t, r, \phi;
\\
i,j,k,\dots&\in&\{ 0,1\} \hbox{ corresponding to } t, r.
\end{eqnarray}
Then
\begin{equation}
\d s^2
=  g_{ij}(t,r)  \d x^i \; \d x^j + 2 g_{\phi i}(t,r) \;\d\phi\;\d x^i + g_{\phi\phi}(t,r) \d\phi^2.
\end{equation}
Furthermore, without any (significant) loss of generality we can choose our $r$ coordinate to satisfy $g_{\phi\phi}(r,t) = r^2$. This is equivalent to choosing $\ell = 2\pi r$ to be the length of the orbit of the axial Killing vector, so that $r$ is a good coordinate as long as $\d\ell\neq0$.  A similar technical restriction applies to the choice of Schwarzschild $r$ coordinate in spherical symmetry. Note that it is this $\d \ell\neq 0$ condition that fails for the self-dual Coussaert--Henneaux spacetimes considered in~\cite{AMZ};  in those spacetimes one instead has $g_{\phi\phi}\to {1\over 4|\Lambda|}$, and the spatial slices are cylindrical rather  than being even approximately planar. This is a situation that is of no direct relevance to stellar structure, in any number of dimensions.  With our preferred choice of $r$ coordinate:
\begin{equation}
\d s^2 =  g_{ij}(t,r)  \d x^i \; \d x^j + 2 g_{\phi i}(t,r) \;\d\phi\;\d x^i + r^2 \d\phi^2.
\end{equation}
This form of the metric still contains \emph{five} free functions of ($r,t$), and is invariant under any coordinate transformation of the form:
\begin{eqnarray}
t&\to& t + f(r,t),\\
r &\to& r,\\
\phi &\to& \phi + h(r,t).
\end{eqnarray}
Under these transformations
\begin{eqnarray}
\d\phi &\to& \d \phi + h_{,r} \d r + h_{,t} \d t,
\\
\d t &\to& \d t + f_{,r} \d r + f_{,t} \d t.
\end{eqnarray}
We can use this remaining coordinate freedom to eliminate two of the metric components; we find it most useful to eliminate $g_{\phi r}$ and $g_{tr}$. A brief computation yields
\begin{equation}
g_{\phi r} \to g_{\phi r} + g_{\phi t} f_{,r} + r^2 h_{,r} ;
\end{equation}
\begin{eqnarray}
g_{tr} &\to& (1+f_{,t})[g_{tr}  + g_{tt} f_{,r}  +  g_{\phi t}  h_{,r} ]
\nonumber
\\
&&  + h_{,t} [g_{\phi r} +  g_{\phi t}  f_{,r} + r^2 h_{,r} ].
\end{eqnarray}
Thus we can eliminate \emph{both} $g_{\phi r}$ \emph{and} $g_{tr}$ provided we choose
\begin{eqnarray}
&&g_{\phi t} f_{,r} +  r^2 h_{,r}  = - g_{\phi r};
\\
&& g_{tt} f_{,r}  +  g_{\phi t}  h_{,r} = - g_{tr}.
\end{eqnarray}
In ``generic'' regions, where the determinant $r^2 g_{tt}-g_{\phi t}^2$ is non-zero, this can be done provided we choose
\begin{eqnarray}
f_{,r} &=& {g_{\phi t} g_{\phi r} - r^2 g_{tr} \over r^2 g_{tt}-g_{\phi t}^2 };
\\
h_{,r} &=& {g_{tr} g_{\phi t} - g_{\phi r} g_{tt} \over r^2 g_{tt}-g_{\phi t}^2 }.
\end{eqnarray}
That is,
\begin{eqnarray}
f(r,t) &=& \tilde f(t) + \int_{r_*}^r {g_{\phi t} g_{\phi r} - r^2 g_{tr} \over r^2 g_{tt}-g_{\phi t}^2 } \d \tilde r;
\\
h(r,t) &=& \tilde h(t) + \int_{r_*}^r {g_{tr} g_{\phi t} - g_{\phi r} g_{tt} \over r^2 g_{tt}-g_{\phi t}^2 } \d\tilde r.\;
\end{eqnarray}
Here $r_*$ is just some convenient place to start integrating from, while $\tilde f(t)$ and  $\tilde h(t)$ are arbitrary functions of $t$.
Note that the determinant $r^2 g_{tt}-g_{\phi t}^2$ is zero if and only if the constant-$r$ hyper-surface is light-like.  (See~\cite{AMZ} for some discussion on this point.) Physically we expect this to happen, at worst, at isolated values of $r$, in which case we could either restrict attention to the connected subsets of the $r$--$t$ plane where the determinant is non-zero, or more prosaically just integrate across the pole using a principal-part prescription. 
Subject to this coordinate choice we have now, (for some \emph{new} functions $g_{ij}$), reduced the metric to the canonical form:
\begin{eqnarray}
\label{E:reduced}
\d s^2 &=&  \left\{ g_{tt}(r,t) \d t^2 + g_{rr}(r,t) \d r^2\right\}  
\nonumber\\
&&
+ 2  g_{\phi t}(r,t) \;\d\phi\;\d t  + r^2 \d\phi^2.
\end{eqnarray}
This form of the metric still contains \emph{three} free functions of ($r,t$).
Furthermore, this form of the metric is still invariant under the residual coordinate transformations
\begin{eqnarray}
t&\to& t + \tilde f(t),\\
r &\to& r,\\
\phi &\to& \phi + \tilde h(t).
\end{eqnarray}

%\bigskip
%-----------------------------------------------------------------------------
\noindent\emph{Proving time independence:} 
For calculations it is useful to now perform some strategic redefinitions and put the metric (\ref{E:reduced}) in the form:
\begin{eqnarray}
\d s^2 &=&  \left\{ - [B(r,t) N(r,t)^2 - r^2 \omega(r,t)^2 ]  \d t^2 + {\d r^2\over B(r,t)} \right\} 
\nonumber\\
&& + 2  r^2 \omega(r,t) \;\d\phi\;\d t  + r^2 \d\phi^2.
\end{eqnarray}
There is no loss of generality in doing so, and  this will make the algebra simpler.

Note that because the metric has been chosen to have two zero components, $g_{rt}=0=g_{r\phi}$, the vacuum Einstein equations 
\begin{equation}
R_{ab} = 2 \Lambda g_{ab},
\end{equation}
include the two particularly simple equations
\begin{equation}
R_{rt}=0=R_{r\phi}.
\end{equation}
Explicit calculation (using {\sf Maple} or similar symbolic manipulation packages) yields
\begin{equation}
R_{rt} - \omega R_{r\phi} = - {1\over2r} {\partial_t B\over B},
\end{equation}
implying $\partial_t B=0$; thus the function $B(r,t)\to B(r)$ is a function of $r$ only. Once this condition has been imposed, the component $R_{rt}$ reduces to
\begin{equation}
R_{rt} = - {1\over2} \; {\omega r^2}\; {\partial_t \partial_r \omega \; N - \partial_t N\partial_r \omega\over B N^3}  
\propto \partial_t \left({\partial_r \omega \over N}\right).
\end{equation} 
Consequently
\begin{equation}
 \partial_t \left({\partial_r \omega \over N}\right) =0 \qquad\implies\qquad   {\partial_r \omega \over N}= F(r),
\end{equation}
and we see
\begin{equation}
\omega(r,t) = \tilde H(t) + \int_{r_*}^r F(\tilde r) \; N(\tilde r,t)\;  \d \tilde r.
\end{equation}
But $\tilde H$ can be absorbed into a redefinition $\phi \to\phi + \tilde h(t)$, so without any loss of generality
\begin{equation}
\omega(r,t)  =  \int_{r_*}^r F(\tilde r) \; N(\tilde r,t) \; \d \tilde r.
\end{equation}
That is: So far, using only the two field equations $R_{rt}=0=R_{r\phi}$, we have deduced
\begin{equation}
B(r,t) = B(r); \qquad \omega(r,t) =  \int_{r_*}^r F(\tilde r) \; N(\tilde r,t) \; \d \tilde r.
\end{equation}
But now the component $R_{rr}$ reduces to
\begin{equation}
R_{rr} =  - {r^4  F(r)^2  \over2} - r \partial_r B  - r B(r) {\partial_r N(r,t)\over N(r,t)},
\end{equation}
and since the $rr$ component of the Einstein equations is $R_{rr} = 2 \Lambda r^2$, we have
\begin{equation}
 {r^4 B(r) F(r)^2 + r \partial_r B  \over2} +  r B(r) {\partial_r N(r,t)\over N(r,t)} = - 2 \Lambda r^2,
\end{equation}
which can be rewritten in the form
\begin{equation}
{\partial_r N(r,t)\over N(r,t)} = X(r).
\end{equation}
But for each value of $t$ this is a first-order homogeneous differential equation in $N(r,t)$, 
implying
\begin{equation}
N(r,t) = E(t) \; N(r),
\end{equation}
where we do not for current purposes need to know the explicit form of $N(r)$.  But this now implies
\begin{equation}
\omega(r,t) =  E(t) \int_{r_*}^r F(\tilde r) N(\tilde r) \, \d \tilde r = E(t) \; \omega(r).
\end{equation}
What this has done for us, using only three of the vacuum field equations, is to show that:
\begin{eqnarray}
B(r,t) &=& B(r);  
\\
N(r,t) &=& E(t)\; N(r);
\\
\omega(r,t) &=&  E(t) \; \omega(r); 
\end{eqnarray}
which now implies
\begin{eqnarray}
\d s^2 &=&  \left\{- [B(r) N(r)^2 - r^2 \omega(r)^2 ]  E(t)^2 \d t^2 + {\d r^2\over B(r)} \right\}  
\nonumber
\\
&&
+ 2  r^2 \omega(r) E(t) \;\d\phi\;\d t  + r^2 \d\phi^2.
\end{eqnarray}
But if we now introduce a new time variable
\begin{equation}
\d t \to E(t) \d t; \qquad t \to \int E(t) \d t
\end{equation}
then this reduces to
\begin{eqnarray}
\label{E:stationary}
\d s^2 &=&  \left\{-[B(r) N(r)^2 - r^2 \omega(r)^2 ]   \d t^2 + {\d r^2\over B(r)} \right\} 
\nonumber
\\
&& + 2  r^2 \omega(r) \;\d\phi\;\d t  + r^2 \d\phi^2.
\end{eqnarray}
This has now reduced the solution of the (2+1) axially symmetric (and \emph{a priori} possibly time dependent) vacuum Einstein equations into a manifestly time-independent (and therefore stationary) form.

%\bigskip
%--------------------------------------------------------------------------------------------
\noindent\emph{Verifying BTZ:} At this stage, because we have now proved time independence,  the BTZ analysis~\cite{BTZ, BTZ2} takes over --- and the solution in the vacuum region outside the rotating star must be a portion of the BTZ spacetime; the (2+1) analogue of Kerr--de~Sitter/Kerr/Kerr-anti-de~Sitter spacetime. It is worthwhile to present an independent derivation of this result. From the stationary metric (\ref{E:stationary}) it is relatively easy, (using {\sf Maple} or similar symbolic manipulation packages), to calculate
\begin{equation}
R_{tt} + B^2 N^2 R_{rr} + \omega^2 R_{\phi\phi} - 2 \omega R_{\phi t} = {B^2 N \partial_r N\over r},
\end{equation}
and
\begin{equation}
g_{tt} + B^2 N^2 g_{rr} + \omega^2 g_{\phi\phi} - 2 \omega g_{\phi t} = 0.
\end{equation}
So from this particular linear combination of the vacuum Einstein equations, $R_{ab} = 2\Lambda g_{ab}$, we deduce $\partial_r N=0$. Therefore $N(r)$ is in fact a constant; call it $E$.

Subject to $N(r)\to E$, we now compute
\begin{equation}
r^2 B R_{rr} + R_{\phi\phi} = -{r\over2} \left\{ r \partial_r^2 B + 3 \partial_r B \right\},
\end{equation}
and
\begin{equation}
r^2 B g_{rr} + g_{\phi\phi} = 2 r^2.
\end{equation}
The corresponding linear combination of vacuum Einstein equations is now
\begin{equation}
 -{r\over2} \left\{ r \partial_r^2 B + 3 \partial_r B \right\} = 4 \Lambda r^2,
\end{equation}
with explicit solution
\begin{equation}
B(r) = 1 - M - \Lambda r^2 + K^2/r^2,
\end{equation}
where $M$ and $K$ are at this stage just constants of integration.
Substituting this back into the Ricci scalar
\begin{equation}
R = 6 \Lambda - {2K^2\over r^4} + {r^2 (\partial_r \omega)^2\over2 E^2},
\end{equation}
the vacuum Einstein equations then yield a simple differential equation for $\omega(r)$:
\begin{equation}
\partial_r \omega =   \pm {2E K\over r^3},
\end{equation}
with explicit solution
\begin{equation}
\omega(r) = E \left(\Omega_0 \pm {  K\over r^2}\right),
\end{equation}
where $\Omega_0$ is an arbitrary constant of integration.
We have now completely solved the vacuum Einstein equations, and the resulting spacetime metric is
\begin{eqnarray}
\d s^2 \!\! &=& \!\!- \left\{ \left[1-M-\Lambda r^2 + {K^2\over r^2} \right] \!+\! r^2 \!\left[ \Omega_0 \pm {K\over r^2} \right]^2 \right\} E^2 \d t^2
\nonumber
\\
&& +{\d r^2 \over\displaystyle 1-M-\Lambda r^2 + {K^2\over r^2} } + r^2 \d \phi^2 
\nonumber
\\
&&+ 2 r^2 \left( \Omega_0 \pm {K\over r^2} \right) E \; \d\phi \d t.
\end{eqnarray}
It is now clear that the constant $E$ can be absorbed into a redefinition of $t$, where $t \to t/E$. Subsequently, the constant $\Omega_0$ corresponds to a ``rigid rotation'' of the spacetime which can be eliminated by a coordinate transform $\phi\to\phi-\Omega_0 t$. Substituting $K=J/2$, and simplifying (in particular, the $g_{tt}$ component), the metric becomes
\begin{eqnarray}
\d s^2 \!\! &=& \!\!- \left\{ 1-M-\Lambda r^2  \right\} \d t^2 +  J \; \d\phi \d t
\nonumber
\\
&& +{\d r^2 \over \displaystyle 1-M-\Lambda r^2 + {J^2\over 4r^2} } + r^2 \d \phi^2.
\end{eqnarray}
This is the BTZ metric~\cite{BTZ, BTZ2}, modulo their unusual normalization for the mass: $M_\mathrm{BTZ}= M_\mathrm{here}-1$. We prefer a normalization for the mass where it is clear that (2+1) Minkowski space is recovered as $\{M,\Lambda,J\}\to 0$. 
This can also be converted to Painlev\'e--Gullstrand form:
\begin{eqnarray}
\d s^2 &=& -  \d t^2  +\left\{\d r+\sqrt{ M+\Lambda r^2 -  {J^2\over 4r^2}} \d t \right\}^2
\nonumber
\\
&&+ r^2 \left\{\d \phi+{J\over2r^2}\d t\right\}^2.
\end{eqnarray}
Remember that we are not asserting that this metric holds throughout the spacetime; for our purposes this metric is valid only in the vacuum region exterior to the rotating and possibly pulsating (2+1) star.

%\bigskip
%--------------------------------------------------------------------------------------------
\noindent\emph{Discussion:} The fact that (under physically plausible technical restrictions) a Birkhoff-like theorem exists for rotating and possibly pulsating stars in (2+1) dimensions is at first rather unexpected: Such theorems certainly do not exist for rotating stars in (3+1) dimensions or higher. The result in (2+1) dimensions depends crucially on the relative simplicity of the Einstein equations in (2+1) dimensions.  Outside the star the spacetime is in fact a constant curvature spacetime --- either de~Sitter, Minkowski, or anti-de~Sitter) --- the only nontrivial physics comes from the holonomies associated with closed paths that encircle the star. (See, for instance,~\cite{AMZ}.)  The (2+1) Birkhoff-like  result indicates that these holonomies are ``frozen in'' and cannot be affected by stellar pulsations, certainly as long as the (2+1) star remains axially symmetric. 

Physically, we feel that it is the absence of gravitational radiation in (2+1) dimensions that lies at the heart of the (axial)+(vacuum) $\implies$ (stationary) result; without gravitational radiation there is no way for the (2+1) star to communicate information regarding its internal state to the exterior region.  This is effectively a ``no hair theorem'', in (2+1) dimensions for stars rather than black holes --- in axial symmetry  the exterior geometry is completely specified by the mass, angular momentum, and cosmological constant. 

%\bigskip

We have not included either electric or magnetic charges in our analysis. It would also be interesting to seek an interior solution.

%\bigskip
%--------------------------------------------------------------------------------------------
\noindent\emph{Acknowledgments:} This research was supported by the Marsden Fund administered by the Royal Society of New Zealand. JS was also supported by Victoria University of Wellington. We wish to thank Thomas Sotiriou for interesting comments and suggestions, and to thank Julio Eduardo Oliva ZapataÄ for bringing reference~\cite{AMZ} to our attention.

%\bigskip

%------------------------------------------------

%------------------------------------------------

%---------------------------------------------------
\end{document}